\documentstyle[12pt,epsfig,axodraw]{article}

\renewcommand{\thefootnote}{\fnsymbol{footnote}}

\newcommand{\s}{\\ \vspace*{-3mm} }

\begin{document}

\newpage
\setcounter{page}{0}

\begin{titlepage}
\begin{flushright}
\hfill{YUMS 98--12}\\
\hfill{SNUTP 98--084}\\
\hfill{August, 1998}
\end{flushright}
\vspace*{1.0cm}

\begin{center}
{\large \bf Phenomenological Impacts of the CP-odd Rephase-invariant\\
            Phase of the Chargino Mass Matrix in the Production\\
            of Light Chargino-Pair in $e^+e^-$ Collisions}
\end{center}
\vskip 1.cm
\begin{center}
{\sc S.Y. Choi$^1$\footnote{e-mail: sychoi@theory.yonsei.ac.kr}, 
     J.S. Shim$^2$\footnote{e-mail: jsshim@wh.myongji.ac.kr}, 
     H.S. Song$^3$\footnote{e-mail: hssong@physs.snu.ac.kr}} 
   and 
{\sc W.Y. Song$^3$\footnote{e-mail: misson@fire.snu.ac.kr}} 

\vskip 0.8cm

\begin{small} 
$^1$ Department of Physics, Yonsei University, Seoul 120--749, Korea 
\vskip 0.2cm
$^2$ Department of Physics, Myong Ji University, Yongin 449-728, Korea 
\vskip 0.2cm
$^3$ Center for Theoretical Physics and Department of Physics, \\
     Seoul National University, Seoul 151-742, Korea
\vskip 0.2cm
\end{small}
\end{center}

\vskip 3cm

\setcounter{footnote}{0}
\begin{abstract}
One CP--odd rephase-invariant phase appears in the chargino mass matrix 
in the minimal Supersymmetric Standard Model. We investigate in detail
the phenomenological impacts of the CP-odd complex phase in the production 
of light charginos in $e^+e^-$ annihilation. The values of the chargino
masses and the mixing angles, determining the size of the wino and higgsino
components in the chargino wave functions, are so sensitive to the CP-odd 
phase that the constraints on the supersymmetric parameters based on the 
conventional assumptions for the parameters are recommended to be 
re-evaluated including the CP-odd phase.
\end{abstract}
\end{titlepage}

\newpage
\renewcommand{\thefootnote}{\alph{footnote}}

\subsection*{1. Introduction}
\label{sec: introduction}

The Minimal Supersymmetric Standard Model (MSSM) \cite{R1} is a well-defined 
quantum theory of which the Lagrangian form is completely known, including 
the general R-parity preserving, soft supersymmetry (SUSY) breaking
terms. 

The full MSSM Lagrangian has 124 truly independent parameters - 79 real 
parameters and 45 CP-violating complex phases \cite{Sutter}. 
The number of parameters in MSSM is too large compared to 19 in the 
Standard Model (SM). Therefore, many studies on possible direct and 
indirect SUSY effects have been made by making several assumptions and 
investigating the variation of a few parameters \cite{R6,Choi2,R6A}.
Recently, it has, however, been shown \cite{Kane} that limits on sparticle
masses and couplings are very sensitive to the assumptions and need to be 
re-evaluated without making any of the simplifying assumptions that have 
been standard. 

Despite the large number of phases in the model as a whole, just one 
CP-odd rephase-invariant phase \cite{Oshimo}, stemed from the chargino 
mass matrix, takes part in chargino production \cite{Ellis}. 
In light of this aspect, in the chargino system the analyses with the 
general parameter set are not so much  more difficult than those with 
parameters assumed real.
The CP-odd phase may be constrained indirectly by the electron or neutron 
electric dipole moment \cite{Oshimo} and may be small, but the constraints
on its actual size depends strongly on the assumptions in those
analyses.  As a matter of fact, the analysis of the electric dipole 
moments has not been done with a general phase structure, which will
certainly give weaker constraints on the CP-odd phase in the 
chargino mass matrix. So, unless there exist any concrete 
demonstrations for a small value of the phase, it will be more
reasonable to take the CP-odd phase as a free parameter.

Charginos are produced in $e^+e^-$ collisions, either in diagonal
or in mixed pairs. However, the second chargino $\tilde{\chi}_2^\pm$ is 
generally expected to be significantly heavier than the first state. 
At LEP2 \cite{R3}, and potentially even in the first phase of $e^+e^-$ 
linear colliders (see e.g. Ref.~\cite{R4}), the chargino 
$\tilde{\chi}_1^\pm$ may  be, for some time, the only chargino state 
that can be studied experimentally in  detail. 
In the present note, we will focus on the diagonal pair production of 
the light chargino $\tilde{\chi}^\pm_1$ in $e^+e^-$ collisions,
\begin{eqnarray*}
e^+e^-\rightarrow \tilde{\chi}_1^+\tilde{\chi}_1^-
\end{eqnarray*}
and investigate in detail the phenomenological impacts of the CP-odd 
complex phase in the determinations of the relevant SUSY parameters 
in the production process.

The production of the light chargino-pair is completely described 
by the chargino mass $\tilde{\chi}^\pm_1$, two mixing angles determining
the size of the wino and higgsino components in the charginos, and the
sneutrino mass. So, first of all, in Section~2, we briefly recapitulate
the elements of the mixing formalism and quantitatively discuss the 
dependence of the chargino masses and the mixing angles on the CP-odd
phase. In Section~3 the cross section for chargino production along with
the light chargino mass is mapped over the parameter space, especially
for the gaugino mass $M_2$ and the higgsino mass parameter $|\mu|$,
by varing the CP-odd phase. Then, we examine the phenomenological impacts 
of the CP-odd phase in constraining the parameter space. 
Conclusions are given in Section~4.

\subsection*{2. Chargino Masses and Mixing Angles}
\label{sec: masses and mixing angles}

In the MSSM, the spin--1/2 partners of the $W$ boson and 
charged Higgs boson, $\tilde{W}^\pm$ and $\tilde{H}^\pm$, mix to form 
chargino mass eigenstates $\tilde{\chi}^\pm_{1,2}$. The mass eigenvalues 
$m_{\tilde{\chi}_{1,2}^\pm}$ and the mixing angles and phases
are determined by the elements of the chargino mass matrix in the
$(\tilde{W}^-,\tilde{H}^-)$ basis \cite{R1} 
\begin{eqnarray}
{\cal M}_C=\left(\begin{array}{cc}
                M_2                &      \sqrt{2}m_W\cos\beta  \\
             \sqrt{2}m_W\sin\beta  &             \mu   
                  \end{array}\right),
\label{eq:mass matrix}
\end{eqnarray}
which is built up by the fundamental supersymmetric parameters;
the gaugino mass $M_2$, the Higgs mass parameter $\mu$, and the ratio 
$\tan\beta=v_2/v_1$ of the vacuum expectation values of the two neutral 
Higgs fields which break the electroweak symmetry. In CP--noninvariant 
theories, the gaugino mass $M_2$ and the Higgs mass parameter $\mu$ 
can be complex. However, by reparametrizations of the fields, $M_2$ 
can be assumed real and positive without loss of generality \cite{Oshimo} 
so that the only non--trivial invariant phase is attributed to $\mu$:
\begin{eqnarray}
\mu = |\mu| {\rm e}^{i\theta_\mu}.
\end{eqnarray}
In these theories the complex chargino mass matrix (\ref{eq:mass matrix})
is diagonalized by two unitary matrices $U_L$ and $U_R$, which can be
parameterized in the following way: 
\begin{eqnarray}
&& U_L=
      \left(\begin{array}{cc}
     \cos\phi_L                     &  {\rm e}^{-i\beta_L}\sin\phi_L \\
      -{\rm e}^{i\beta_L}\sin\phi_L  &  \cos\phi_L
      \end{array}\right), \nonumber \\    
&& U_R=\left(\begin{array}{cc}
     {\rm e}^{i\gamma_1}   &    0   \\
                   0       &  {\rm e}^{i\gamma_2}
             \end{array}\right)
      \left(\begin{array}{cc}
     \cos\phi_R                     &  {\rm e}^{-i\beta_R}\sin\phi_R \\
      -{\rm e}^{i\beta_R}\sin\phi_R  &  \cos\phi_R
      \end{array}\right),  
\label{eq:mixing matrix}
\end{eqnarray}
and which render $U_R{\cal M}_C U^\dagger_L$ diagonal.
The two chargino mass eigenvalues are given by
\begin{eqnarray}
m^2_{\tilde{\chi}^\pm_{1,2}}=\frac{1}{2}\bigg[M_2^2+|\mu|^2+2m^2_W
      \mp\Delta_C\bigg],
\label{eq:mass}
\end{eqnarray}
with 
\begin{eqnarray}
\Delta_C= \sqrt{(M_2^2-|\mu|^2)^2+4m^4_W\cos^2 2\beta+4m^2_W(M_2^2+|\mu|^2
         +2M_2|\mu|\sin 2\beta\cos\theta_\mu)}.
\end{eqnarray}
The mixing angles $\phi_{L,R}$ are given by the relations
\begin{eqnarray}
&&\cos 2\phi_L=-\frac{M_2^2-|\mu|^2-2m^2_W\cos 2\beta}{\Delta_C},
              \nonumber\\ 
&&\sin 2\phi_L=-\frac{2m_W\sqrt{M^2_2+|\mu^2|+(M_2^2-|\mu|^2)\cos 2\beta
                     +2M_2|\mu|\sin 2\beta\cos\theta_\mu}}{\Delta_C},
\label{eq:phi-L}
\end{eqnarray}
\begin{eqnarray}
&&\cos 2\phi_R=-\frac{M_2^2-|\mu|^2+2m^2_W\cos 2\beta}{\Delta_C},
              \nonumber\\
&&\sin 2\phi_R=-\frac{2m_W\sqrt{M^2_2+|\mu^2|-(M_2^2-|\mu|^2)\cos 2\beta
                     +2M_2|\mu|\sin 2\beta\cos\theta_\mu}}{\Delta_C},
\label{eq:phi-R}
\end{eqnarray}
The four nontrivial phase angles $\{\beta_L,\beta_R,\gamma_1,\gamma_2\}$ 
also depend on the invariant angle $\theta$; for their expressions  
we refer to the appendix of the work \cite{Choi2}.

Note that $\cos\theta_\mu$ in eqs.~(\ref{eq:mass}), (\ref{eq:phi-L})
and (\ref{eq:phi-R}) for the chargino masses and the mixing angles appears
along with a unique combination factor:
\begin{eqnarray*}
M_2|\mu|\sin 2\beta=M_2|\mu|\:\frac{2\tan\beta}{1+\tan^2\beta}.
\end{eqnarray*}
This is a reflection of the fact that the CP-odd phase angle $\theta_\mu$ 
can be absorbed by field re--definitions if at least one of the chargino 
mass matrix elements vanishes. In particular, when $\tan\beta$ is 
very small or very large, i.e. one of the two Higgs vacuum expectation 
values $v_1$ and $v_2$ is relatively very small \footnote{Two 
vacuum expectation values $v_1$ and $v_2$ can not be larger than 
$v=\sqrt{v^2_1+v^2_2}\approx 250$ GeV.}, 
the effects of the phase angle $\theta_\mu$ diminish. 
Keeping in mind that the CP-odd phase effects are very small for
large $\tan\beta$, we present numerical analyses for a fixed value of
$\tan\beta=2$ in the following.

The light and heavy chargino masses are presented in Figs.~1(a) and (b) 
as a function of the cosine of the CP-violating phase angle $\theta_\mu$ for a 
representative set of parameters. 
The parameters are chosen in the higgsino region $M_2 \gg |\mu|$, the gaugino 
region $M_2 \ll |\mu|$ and in the mixed region $M_2 \sim |\mu|$ for 
$\tan\beta=2$ as
\begin{eqnarray}
\begin{array}{ll}
{\rm gaugino\ \ region} \makebox[1mm]{}:
 & (M_2,|\mu|)=(80\ \ {\rm GeV}, \ \ 200\ \ {\rm GeV}),\\ 
{\rm higgsino\ \ region}: 
 & (M_2,|\mu|)=(210\: {\rm GeV},\ \ \, 70\ \ {\rm GeV}),\\ 
{\rm mixed\ \ region} 
  \makebox[4mm]{}: 
 & (M_2,|\mu|)=(90\ \ {\rm GeV},\ \ \ 90\ \ {\rm GeV}).
\label{eq:parameter}
\end{array}
\end{eqnarray}
The two masses are very sensitive to the phase angle $\theta_\mu$ in all
scenarios; the sensitivity is more prominent in the light chargino 
mass than in the heavy chargino mass,  and it is most prominent
in the mixed scenario. Figs.~1(c) and (d) exhibit $\cos 2\phi_L$ and 
$\cos 2\phi_R$ as a function of $\cos\theta_\mu$\footnote{The two sines,
$\{\sin 2\phi_L$,$\sin 2\phi_R\}$, and the four nontrivial phase angles 
$\{\beta_{L,R},\gamma_{1,2}\}$ are not numerically presented in the present
work because they are not involved in the diagonal pair production of the 
light or heavy charginos as shown in the next Section.}. 
Similarly, both of them depend more strongly on the CP-violating phase 
angle $\theta_\mu$ in the mixed scenario than in the gaugino and higgsino 
scenarios.

\subsection*{3. Production Cross Section}
\label{sec: production x-section}

The process $e^+e^-\rightarrow\tilde{\chi}^+_1\tilde{\chi}^-_1$ is generated
by the three mechanisms: $s$--channel $\gamma$ and $Z$ exchanges, and 
$t$--channel $\tilde{\nu}$ exchange. The transition matrix element, 
after a Fierz transformation of the $\tilde{\nu}$--exchange amplitude,
\begin{eqnarray}
T\left(e^+e^-\rightarrow\tilde{\chi}^+_1\tilde{\chi}^-_1\right)
 = \frac{e^2}{s}Q_{\alpha\beta}
   \left[\bar{v}(e^+)  \gamma_\mu P_\alpha  u(e^-)\right]
   \left[\bar{u}(\tilde{\chi}^-_1) \gamma^\mu P_\beta 
               v(\tilde{\chi}^+_1)\right],
\label{eq:production amplitude}
\end{eqnarray}
can be expressed in terms of four bilinear charges, classified according 
to the chiralities $\alpha,\beta=L,R$ of the associated lepton and 
chargino currents
\begin{eqnarray}
Q_{LL}&=&1+ \frac{D_Z}{s_W^2 c_W^2}(s_W^2 -\frac{1}{2}) 
         \left(s_W^2 -\frac{3}{4}-\frac{1}{4}\cos 2\phi_L\right),
         \nonumber\\ 
Q_{LR}&=&1+ \frac{D_Z}{s_W^2 c_W^2} (s_W^2 -\frac{1}{2}) 
         \left(s_W^2-\frac{3}{4}-\frac{1}{4}\cos 2\phi_R\right) 
        + \frac{D_{\tilde{\nu}}}{4s_W^2} (1+\cos 2\phi_R), \nonumber\\
Q_{RL}&=&1+\frac{D_Z}{c_W^2} \left(s_W^2 -\frac{3}{4}-\frac{1}{4}\cos 
          2\phi_L\right), \nonumber\\
Q_{RR}&=&1+ \frac{D_Z}{c_W^2}  \left(s_W^2 -\frac{3}{4}-\frac{1}{4}\cos 
         2\phi_R\right).
\end{eqnarray}
The first index in $Q_{\alpha \beta}$ refers to the chirality of the $e^\pm$ 
current, the second index to the chirality of the $\tilde{\chi}_1^\pm$ current.
The $\tilde{\nu}$ exchange affects only the $LR$ chirality charge while all 
other amplitudes are built up by $\gamma$ and $Z$ exchanges. $D_{\tilde{\nu}}$ 
denotes the sneutrino propagator $D_{\tilde{\nu}} = s/(t- m_{\tilde{\nu}}^2)$, 
and $D_Z$ the $Z$ propagator $D_Z=s/(s-m^2_Z+im_Z\Gamma_Z)$; 
the non--zero width can in general be neglected for the energies considered 
in the present analysis so that the charges are real. \s 

The bilinear charges $\{Q_{RR},Q_{RL}\}$ ($\{Q_{LR},Q_{LL}\})$
with right (left) electron chirality depend only on $\cos 2\phi_R$ 
($\cos 2\phi_L$) so that the dependence of the production amplitude
on the CP-odd phase angle $\theta_\mu$ is easily readable. 
Fig.~2 shows the total production cross section as a function 
of (a) $\cos\theta_\mu$ for a fixed sneutrino mass of 200 GeV and the
parameter in eq.~(\ref{eq:parameter}), and (b) the sneutrino 
mass in the mixed scenario of the parameter set (\ref{eq:parameter})
at a c.m. energy of 200 GeV.
Interestingly, while the chargino masses and the mixing angles 
are most sensitive to the CP-violating phase angle in the mixed scenario, 
the production cross section itself is most sensitive to the phase in
the gaugino scenario. This implies that the sensitivities to the phase 
angle $\theta_\mu$ is strongly correlated with those to the sneutrino mass 
involved in the $t$-channel snuetrino exchange amplitude. 
Fig.~2(b) clearly shows that the total cross section grows up very 
sharply with $\cos\theta_\mu$ and the sneutrino mass for largest destructive 
interference in the production amplitude shifts to a lower value. 
Prior or simultaneous determination of $m_{\tilde{\nu}}$ will be therefore 
necessary to determine the phase and the other SUSY parameters. \s

The maximal $(M_2,|\mu|)$ parameter space, probed in the production of
light charginos in $e^+e^-$ collisions with a c.m. energy of 200 GeV,
is presented in Fig.~3(a) for three different values of $\cos\theta_\mu$ 
with $\tan\beta=2$. 
As $\cos\theta_\mu$ increases, the covered parameter space is 
enlarged. However, the maximal parameter space can not be fully covered 
in actual experiments, but its searchable regime relies  
on the number of produced charginos and the reconstruction efficiency 
of the chargino signals determined from their decay patterns \cite{PDG98}. 
In the case that the light chargino decays into 
the lightest neutralino, [usually considered to be the lightest 
supersymmetric particle (LSP)], a fermion and an anti-fermion, 
the difference between the chargino mass and the LSP mass plays 
a crucial role. Since the LSP mass also depends strongly on the 
CP-odd phase angle $\theta_\mu$ as well as an extra phase 
angle \cite{Kane}, the determination of the upper limits of 
the chargino cross section, which can be excluded at high energy 
colliders such as LEPII, will be rather involved, but of course 
doable. 
In the present work, we simply take $\sigma_{tot}=2$ pb as a 
reference value for the limit of the production cross section while
its estimates based on the detailed investigations of the chargino
decays will be touched on in our future work. Fig.~3(b) shows the
contours for the total cross section $\sigma_{tot}= 2$ with 
$m_{\tilde{\nu}}=200$ GeV at a c.m. energy of 200 GeV for three
different values of $\cos\theta_\mu$. It is clear that the excluded
region of the parameter space depend strongly on the phase angle 
$\theta_\mu$ and
its dependence is most prominent in the mixed scenario.

\subsection*{4. Conclusions}
\label{sec: conclusions}

We have analyzed how the parameters of the chargino system, the masses 
of the charginos $\tilde{\chi}^\pm$ and the size of the wino and 
higgsino components in the chargino wave functions, parametrized
in terms of the two angles $\phi_L$ and $\phi_R$, are affected by
the CP-violating rephase-invariant angle $\theta_\mu$ in the
chargino mass matrix. In addition, we have studied the dependence
of the production cross section of the light chargino-pair in
$e^+e^-$ collisions on the angle $\theta_\mu$ in detail.

The chargino masses and the production cross section of the
light chargino-pair in $e^+e^-$ collisions are very sensitive 
to the CP-violating angle $\theta_\mu$. The sensitivities are most 
prominent around $\tan\beta=1$ and in the mixed scenario with 
$M_2\sim |\mu|$, where the limits on the light chargino mass 
might be much weaker than has been reported \cite{PDG98}
and need to be re-evaluated.
It is then essential to probe the parameter space including the possible
complex phase angle $\theta_\mu$ in the search of the SUSY particles 
at high energy colliders.

\subsubsection*{Acknowledgments}

This work was supported by Korean Science and Engineering Foundation
in part through large collaboration project, Project  No. 96-0702-01-01-2 
and in part through the Center for Theoretical physics, Seoul National 
University.

\bigskip \bigskip

\newpage
\mbox{ }
\vskip  1cm
\begin{center}
\begin{figure}[htb]
\hbox to\textwidth{\hss\epsfig{file=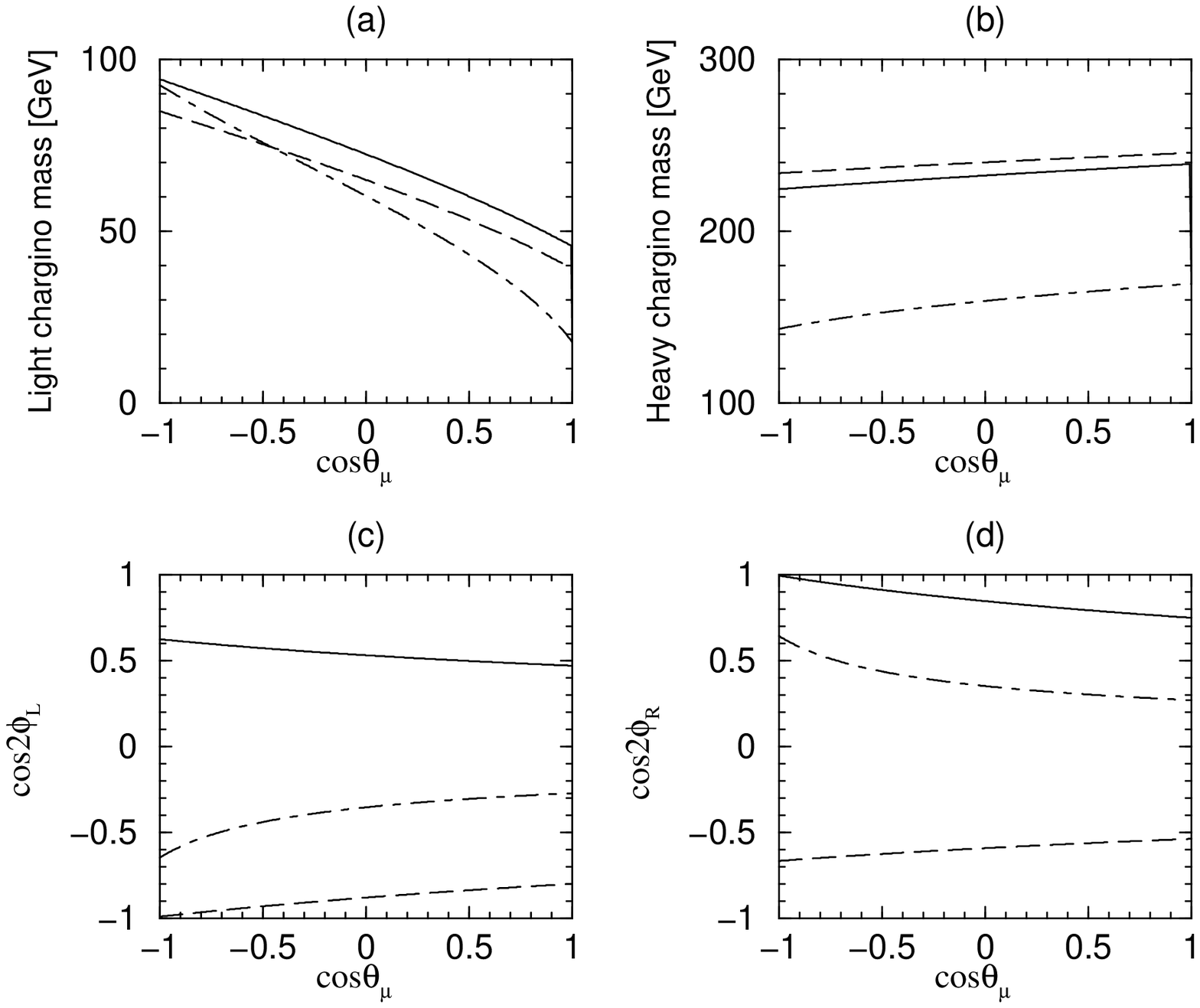,height=14cm}\hss}
\caption{\it (a) the light chargino mass, (b) the heavy chargino mass,
             (c) $\cos 2\phi_L$, and (d) $\cos 2\phi_R$ as a function of
             the cosine of the CP-violating phase angle $\theta_\mu$ for the 
             representative set of SUSY parameters in 
             eq.~(\ref{eq:parameter}): 
             solid line  for the gaugino case, dashed line for the
             higgsino case, and  dot-dashed line for the mixed case.}
\label{fig:cth}
\end{figure}
\end{center}

\vskip -1cm

\newpage
\mbox{ }
\vskip 5cm

\begin{center}
\begin{figure}[htb]
\hbox to\textwidth{\hss\epsfig{file=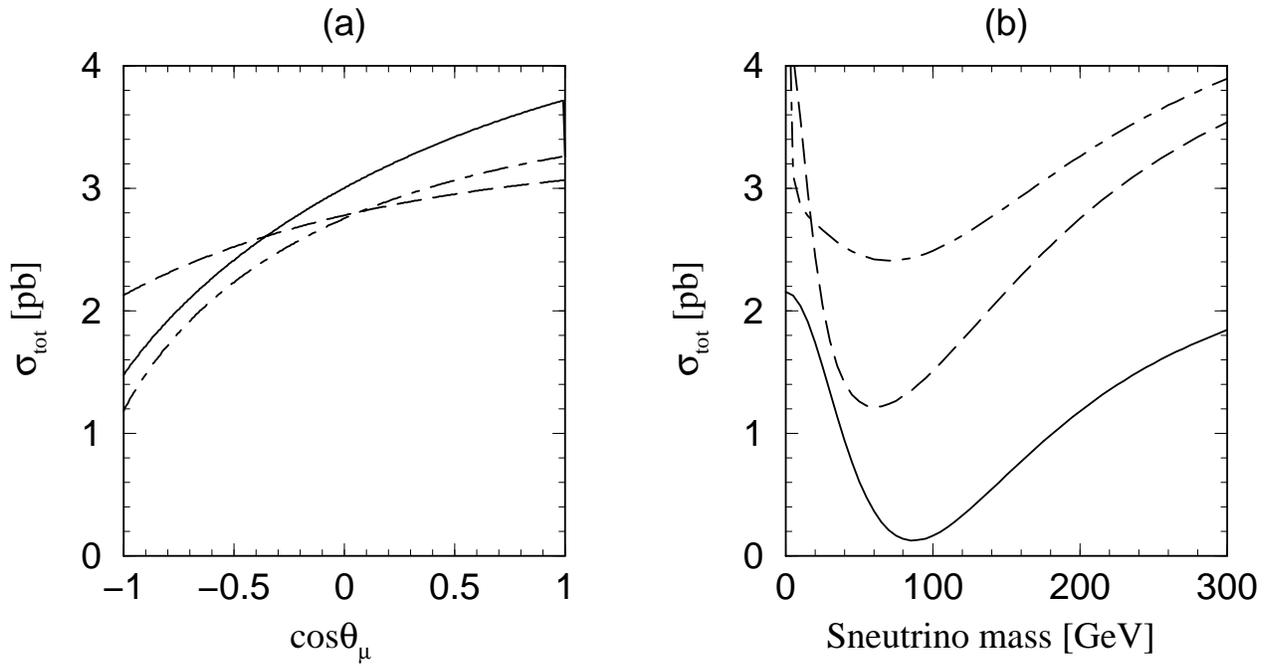,height=9cm}\hss}
\caption{\it The cross section for the production of light charginos 
             (a) as a function of $\cos\theta_\mu$ with 
             $m_{\tilde{\nu}}=200$ GeV for the parameters set in   
             eq.~(\ref{eq:parameter}), and (b) as a function of
             the sneutrino mass in the mixed scenario at 
             $\sqrt{s}= 200$ GeV; in (a) 
             the solid line for the gaugino case, the dashed line for 
             the higgsino case, and the  dot-dashed line 
             for the mixed case, and in (b) the solid line for 
             $\cos\theta_\mu=-1$, the dashed line for $\cos\theta_\mu=0$,
             and the dot-dashed line for $\cos\theta_\mu=1$.}
\label{fig:xrs_th}
\end{figure}
\end{center}
\begin{center}
\begin{figure}[htb]
\hbox to\textwidth{\hss\epsfig{file=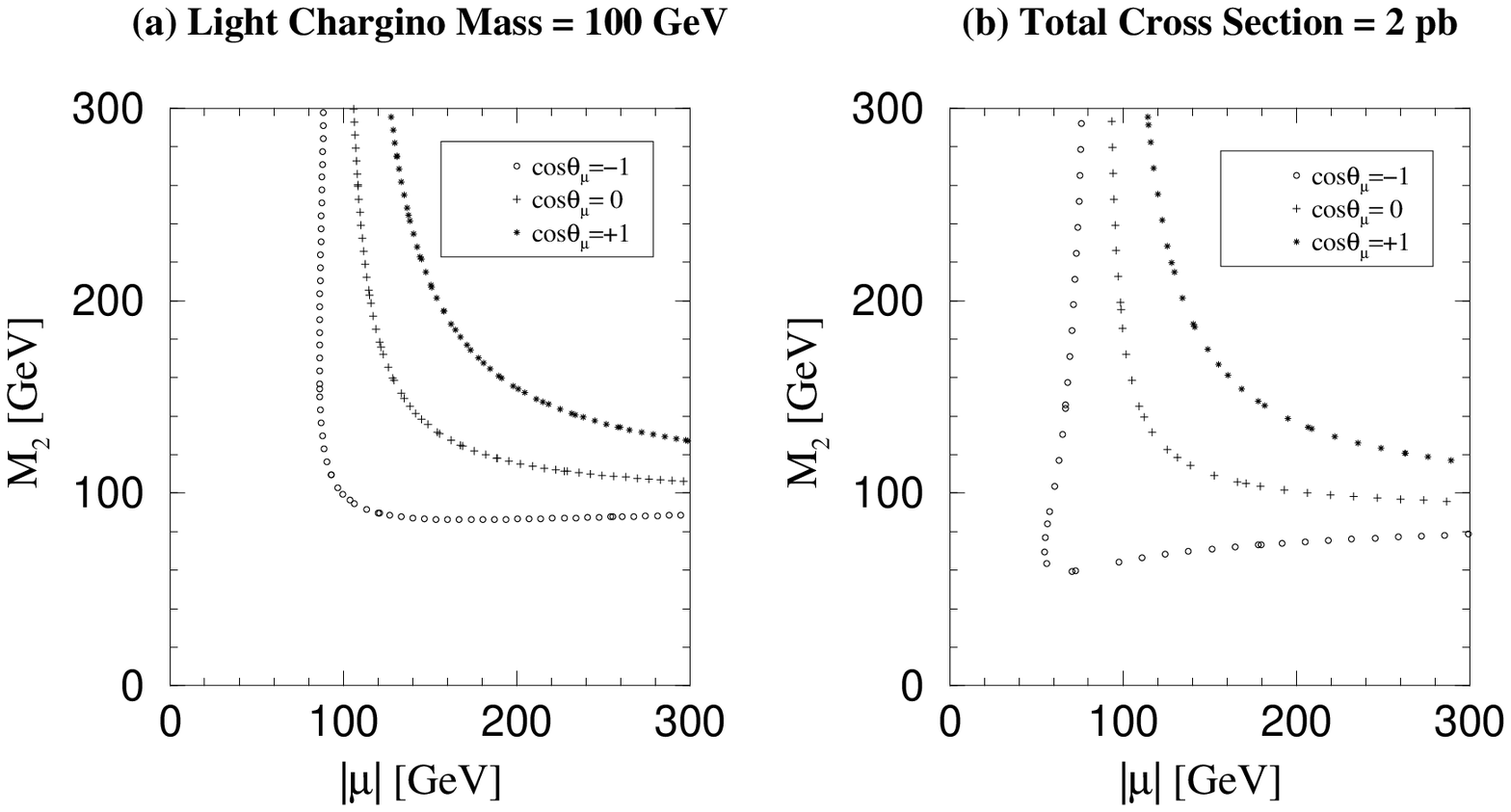,height=9cm}\hss}
\caption{\it Contours for (a) the light chargino mass 
             $m_{\tilde{\chi}^\pm_1}=100$ GeV and (b) the total
             cross section $\sigma_{tot}=2$ pb with 
             $m_{\tilde{\nu}}= 200 $ GeV at $\sqrt{s}=200$ GeV; the line with
             circle symbols for $\cos\theta_\mu=-1$, the line with plus
             symbols for $\cos\theta_\mu=0$ and the line with star symbols
             for $\cos\theta_\mu=+1$.}
\label{fig:xrs_cont}
\end{figure}
\end{center}


\begin{thebibliography}{99}

\bibitem{R1} For reviews of supersymmetry and the Minimal Supersymmetric 
   Standard Model, see H.~Nilles, Phys.~Rep.~{\bf 110} (1984) 1; 
   H.E.~Haber and G.L.~Kane, Phys.~Rep.~{\bf 117} (1985) 75. 

\bibitem{Sutter} S.~Dimopoulos and D.~Sutter, Nucl.~Phys.~{\bf B 452}
   (1995) 496; H.~Haber, Proceedings of the 5th International 
   Conference on Supersymmetries in Physics (SUSY'97), May 1997,
   ed. M.~Cveti\'{c} and P.~Langacker, hep-ph/9709450.

\bibitem{R6} A.~Leike, Int.~J.~Mod.~Phys.~{\bf A3} (1988) 2895; 
   M.A.~Diaz and S.F.~King, Phys.~Lett.~{\bf B349} (1995) 105; {\bf B373} 
   (1996) 100;  J.L.~Feng and M.J.~Strassler, Phys.~Rev.~{\bf D51}
   (1995) 4461 and {\bf D55} (1997) 1326;  G. Moortgat-Pick and
   H.~Fraas, hep-ph/9708481. 

\bibitem{Choi2} S.Y.~Choi {\it et al.},  hep-ph/9806279.

\bibitem{R6A} G.~Moortgat-Pick, H.~Fraas, A.~Bartl and, and W.~Majerotto,
   hep-ph/9804306.

\bibitem{Kane} M.~Brhlik and G.L.~Kane, hep-ph/9803391. 

\bibitem{Oshimo} Y.~Kizukuri and N.~Oshimo, Proceedings of the Workshop on    
   {\it $e^+e^-$ Collisions at 500 GeV: The Physics Potential}, 
   Munich-Annecy-Hamburg 1991/93, DES 92-123A+B, 93-123C, ed. P.~Zerwas;
   T.~Ibrahim and P.~Nath, Phys.~Lett.~{\bf B418} (1998) 98;
   Phys.~Rev.~{\bf D57} (1998) 478; hep-ph/9807591.

\bibitem{Ellis} J.~Ellis, J.~Hagelin, D.~Nanopoulos and M.~Srednicki,
   Phys.~Lett.~{\bf 127B} (1983) 233; V.~Barger, R.W.~Robinett, W.Y.~Keung 
   and R.J.N.~Phillips, Phys.~Lett.~{\bf B131} (1983) 372; D.~Dicuss, 
   S.~Nandi, W.~Repko and X.~Tata, Phys.~Rev.~Lett. {\bf 51} (1983) 1030; 
   S.~Dawson, E.~Eichten and C.~Quigg, Phys.~Rev.~{\bf D31} (1985) 1581; 
   A.~Bartl and H.~Fraas and W.~Majerotto, Z.~Phys.~{\bf C30} (1986) 441.  

\bibitem{R3} Proceedings of the Workshop on 
   {\it Physics at LEP II}, Report No. CERN-96-01, eds. G.~Altarelli, 
   T.~Sj\"{o}strand, and F.~Zwirner.

\bibitem{R4} E.~Accomando {\it et al.}, Phys.~Rept.~{\bf 299} (1998) 1.  

\bibitem{PDG98} See, for example, Particle Data Group, 
   Eur.~Phys.~J.~{\bf C3} (1998) 1.

\end{thebibliography}
\end{document}